\documentclass[aip,jcp,amsmath,amssymb,preprint]{revtex4-1}

\usepackage[version=3]{mhchem}
\usepackage{dcolumn} 
\usepackage{bm}
\usepackage{graphicx}
\usepackage{longtable}

\newcommand*\latin[1]{\textit{#1}}
\newcommand*\ie{\latin{i.e.}}
\newcommand*\abinitio{\latin{ab initio}}
\newcommand*\wn{cm$^{-1}$}
\newcommand*\prog[1]{\texttt{#1}}

\usepackage{color}
\newcommand{\red}[1]{{\color{black} #1}}

\begin{document}

\title{Structure-based Sampling and Self-correcting Machine Learning for 
Accurate Calculations of Potential Energy Surfaces and Vibrational Levels}

\author{Pavlo O. Dral}
\email{dral@kofo.mpg.de}
\affiliation{Max-Planck-Institut f\"ur Kohlenforschung, Kaiser-Wilhelm-Platz 1, 
45470 M\"ulheim an der Ruhr, Germany}

\author{Alec Owens}
\affiliation{Department of Physics and Astronomy, University College London, 
Gower Street, WC1E 6BT London, United Kingdom}
\affiliation{Max-Planck-Institut f\"ur Kohlenforschung, Kaiser-Wilhelm-Platz 1, 
45470 M\"ulheim an der Ruhr, Germany}

\author{Sergei N. Yurchenko}
\affiliation{Department of Physics and Astronomy, University College London, 
Gower Street, WC1E 6BT London, United Kingdom}

\author{Walter Thiel}
\affiliation{Max-Planck-Institut f\"ur Kohlenforschung, Kaiser-Wilhelm-Platz 1, 
45470 M\"ulheim an der Ruhr, Germany}

\date{\today}

\begin{abstract}
We present an efficient approach for generating highly accurate molecular 
potential energy surfaces (PESs) using self-correcting, kernel ridge regression 
(KRR) based machine learning (ML). We introduce structure-based sampling to 
automatically assign nuclear configurations from a pre-defined grid to the 
training and prediction sets, respectively. Accurate high-level \abinitio{} 
energies are required only for the points in the training set, while 
the energies for the remaining points are provided by the ML model with 
negligible computational cost. The proposed sampling procedure is shown to be 
superior to random sampling and also eliminates the need for training several 
ML models. Self-correcting machine learning has been implemented such that each 
additional layer corrects errors from the previous layer. The performance of 
our approach is demonstrated in a case study on a published high-level 
\abinitio{} PES of methyl chloride with 44,819 points. The ML model is 
trained on sets of different 
size and then used to predict the energies for tens of thousands of nuclear 
configurations within seconds. The resulting datasets are utilized in 
variational calculations of the vibrational energy levels of CH$_3$Cl. By using 
both structure-based sampling and self-correction, the size of the training set 
can be kept small (e.g. 10\% of the points) without any significant loss of 
accuracy. In \abinitio{} rovibrational spectroscopy, it is thus possible 
to \red{reduce} the number of computationally costly electronic structure 
calculations through \red{structure-based sampling and} self-correcting KKR-based machine learning
\red{by up to 90\%}.
\end{abstract}

\pacs{}

\maketitle 

\section{Introduction}

 Molecular potential energy surfaces (PESs) are often computed using high-level 
\abinitio{} theory, notably in theoretical ro-vibrational spectroscopy 
where sub-wavenumber accuracy is targeted for calculated transition 
frequencies. Such sophisticated quantum mechanical (QM) calculations can be 
extremely costly, and they have to be repeated for tens of thousands of nuclear 
geometries. Particularly challenging are larger systems with many degrees of 
freedom that suffer from the so-called ``curse of dimensionality''. Machine 
learning (ML) techniques offer a means to significantly reduce the necessary 
computational time~\cite{Behler_Perspective-ML-potentials_2016}. 
 
 It has been shown that neural networks (NNs) can be used to interpolate 
between points along the coordinate of simple 
reactions~\cite{Gastegger_NN_Reaction_2015,Kolb_NN-reaction-PES_2016}, to 
improve the accuracy of semiempirical QM/MM 
calculations\cite{Yang_QMMM-NN_2016}, and to calculate accurate energies of 
conformations and clusters of organic and inorganic 
species~\cite{Bholoa_NN_2007,Malsche_NNPES_2009,Behler_NNmetal_2012,
Schuett_DTNN_NatComm_2017,Isayev_ANI-1_ChemSci_2017}. 
Kernel ridge regression (KRR) can reduce errors of low-level QM methods such 
that atomization enthalpies of organic isomers can be reproduced with an 
accuracy close to higher-level QM 
methods~\cite{Dral_PL_2015,Dral_Delta_ML_2015}.
Gaussian process models are able to sample low-energy regions of the 
PES~\cite{Toyoura_ML_lowPES_2016}, correct density functional theory energies 
of water~\cite{Bartok_GPwater_2013}, predict energies and forces in Si$_n$ 
clusters~\cite{Bartok_SOAP_2013}, and construct global PESs for 
N$_4$~\cite{GP_Cui2016}.
ML can also be employed for accurate yet computationally fast PES 
representation; for example, NNs have been used to compute molecular 
vibrational energy 
levels~\cite{Manzhos_RSHDMRNN_2006,Manzhos_nestedNN_2006,Manzhos_NN_2014,
Majumder_MP_2015,Zhang_FINN_2016}, reaction
probabilities~\cite{Zhang_FINN_2016}, and 
reaction rates~\cite{Doren_NNMC_1995}.
Whilst potentials based on KRR~\cite{Chmiela_KRR-FF_2016}, 
NNs~\cite{Hobday_NN_1999,Hobday_GANN_1999,Raff_MDsamplingNN_2005,
Behler_Parrinello_NN_2007,Behler_SiNN_2008,Pukrittayakamee_NN-PES-FF_2009,
Behler_NaNN_2010,Behler_CarbonNN_2010,
Behler_PCCP_2011,Behler_NatMater_2011,Behler_WaterNN_2012,
Behler_IJQC_NNtutorial_2015} or Gaussian process
models~\cite{Bartok_GAP_2010} have been utilized in 
molecular dynamics (MD) simulations.

 ML performs best in the interpolation regime and can fail spectacularly when 
used for extrapolation. For relatively large training sets and high error 
tolerance, as in many chemical applications, training on a random sample of 
data points can give useful 
results~\cite{CMat,MLalgsAssessment,Dral_PL_2015,Dral_Delta_ML_2015,GP_Cui2016}.
 Random sampling is straightforward and has minimal computational cost involved 
but it does not ensure \latin{per se} that ML avoids extrapolation. Thus, it is 
preferable to ensure that ML is applied only for interpolation by using better 
sampling techniques. Even a simple manual choice of critical points on the PES 
can improve the accuracy of ML significantly~\cite{Gastegger_NN_Reaction_2015}. 
 
 Automated sampling is necessary for more complex PESs and several approaches 
have been reported in the literature. When training ML potentials, structures 
have been screened on their input vector elements to see if they are within the 
minimum and maximum values of the training set, which can then be expanded to 
include `outside-the-range' 
structures~\cite{Behler_PCCP_2011,Botu_KRR-adaptive-MD_2014}. NN PESs of simple 
reactions have been re-fitted with points when the initial NN potential has 
encountered problems~\cite{Kolb_NN-reaction-PES_2016}. Similarly, the error of 
the ML potential can be monitored during dynamics and ML can be re-trained 
on-the-fly by including additional points when errors become too 
large~\cite{DeVita_Gaussian-on-the-fly-forces_2015}. In the case of very simple 
systems like \ce{H3}, the training set can be generated on a coarse 
grid~\cite{Kolb_NN-reaction-PES_2016}. 
  
 Because of their inherent flexibility different NNs can give very different 
energies for the same structure, and hence training different NNs on the same 
data can also highlight problem structures that should be incorporated into the 
training set~\cite{Behler_WaterNN_2012,Behler_NNmetal_2012}. Another approach 
exploits the fact that the most important regions of the PES are likely to be 
covered by standard MD simulations. Snapshots from the MD trajectories can thus 
be used in the training set to produce a more robust ML 
PES~\cite{Raff_MDsamplingNN_2005,Pukrittayakamee_NN-PES-FF_2009,
Geiger_NN-polymorphic_2013,Suzuki_ML-forces_2016,Schuett_DTNN_NatComm_2017,
Chmiela_KRR-FF_2016}.
For better sampling of vibrational phase space, previous studies have employed 
stretched bond distances in MD 
simulations~\cite{Pukrittayakamee_NN-PES-FF_2009} or a random displacement to 
sample along the vibrational normal modes~\cite{Isayev_ANI-1_ChemSci_2017}. In 
other work, structures have been chosen from MD trajectories based on various 
clustering approaches, which reduce the redundancies in the NN training 
set~\cite{Aspuru-Guzik_ML-exciton-dynamics_2016}. For calculations of 
rotation-vibration spectra where low-energy regions of the PES are more 
important, probability distribution functions that favor structures near 
equilibrium have been used for pseudo-random 
sampling~\cite{Manzhos_RSHDMRNN_2006,Manzhos_nestedNN_2006,Manzhos_NN_2014}.

 Here we report a new and efficient method for generating highly accurate 
molecular PESs which uses KRR-based machine learning. In our approach, we 
initially define a set of grid points (nuclear configurations) that covers the 
relevant low-energy part of the PES, which can be done using relatively 
inexpensive energy estimates (see Section~\ref{subsec:sampling}). Thereafter, 
our sampling relies on an interpolation between structures rather than on 
energies, \ie{} the sampling procedure itself does not require energy 
calculations or the training of several ML models. We investigate the effect of 
structure-based sampling on the accuracy of the ML model in comparison with 
random sampling. We also explore what error reduction can be achieved through 
ML self-correction. 
 
 To illustrate our KRR-based ML model we utilize a high-level \textit{ab 
initio} PES for methyl chloride (\ce{CH3{}^{35}Cl})~\cite{15OwYuYa.CH3Cl}. This 
PES, denoted CBS-35$^{\,\mathrm{HL}}$, is based on extensive explicitly 
correlated coupled cluster calculations with extrapolation to the complete 
basis set (CBS) limit and incorporates a range of higher-level additive energy 
corrections. These include: core-valence electron correlation, higher-order 
coupled cluster terms, scalar relativistic effects, and diagonal 
Born-Oppenheimer corrections.
A considerable amount of computational time was spent generating this PES. For 
example, a single point of the CBS-35$^{\,\mathrm{HL}}$ PES at the equilibrium 
geometry required 9 separate calculations and a combined total of 26.7 hours on 
a single core of an Intel Xeon E5-2690 v2 $3.0\,$GHz processor. Building a 
reliable PES requires tens of thousands of nuclear geometries for polyatomic 
species and the computational time increases for distorted configurations due 
to slower energy convergence. Efficient ML techniques that can reduce the 
necessary computational effort are therefore highly desirable.
 
 The paper is structured as follows: In Section~\ref{sec:methods} we present 
the KRR-based self-correcting ML model, the molecular descriptor, and the 
sampling procedure. Details on the variational calculation of vibrational 
energy levels are also given. In Section~\ref{sec:results} the ML model is 
evaluated for different sizes of training set, and vibrational energies are 
computed using several ML model PESs. Concluding remarks are offered in 
Section~\ref{sec:conc} and an outlook is provided in Section~\ref{sec:outlook}.

\section{Methods}
\label{sec:methods}

\subsection{Machine Learning}
\label{subsec:ml}

 The chosen ML technique is based on KRR~\cite{Hastie_Book_2009} and is similar 
to the approach used for ML across chemical compound 
space~\cite{CMat,CMCCS,MLalgsAssessment,Dral_Delta_ML_2015,
Ramakrishnan_ML_TDDFT_2015} and for calculating relative stabilities of organic 
isomers~\cite{Dral_Delta_ML_2015,Dral_PL_2015}. Below we give details of the 
KRR approach relevant to this work. All ML calculations were performed using 
our own program package \prog{MLatom}~\cite{MLatom0.9}.
 
 Some property value $Y^{\mathrm{ML}}\left(\mathbf{M}_i\right)$ of a nuclear 
configuration $i$ represented by the molecular descriptor $\mathbf{M}_i$ 
(discussed in Section~\ref{subsec:moldescr}) is calculated using the 
expression~\cite{Hastie_Book_2009},
\begin{equation}
  Y^{\mathrm{ML}}\left(\mathbf{M}_i\right) = \sum_{j=1}^{N_{\mathrm{train}}}{\alpha_j K\left(\mathbf{M}_i,\mathbf{M}_j\right)},\label{eq:KRRest}
\end{equation}
where the sum runs over $N_{\mathrm{train}}$ configurations represented by the 
molecular descriptors $\{\mathbf{M}_j\}$ of the training set. The regression 
coefficient $\alpha_j$ refers to the configuration $\mathbf{M}_j$, and $K$ is 
the kernel. Here we use the Gaussian kernel which is equivalent to the form 
often employed in the literature~\cite{CMat,MLalgsAssessment},
\begin{equation}
  K\left(\mathbf{M}_i,\mathbf{M}_j\right) = \exp\left(-\frac{D\left(\mathbf{M}_i,\mathbf{M}_j\right)^2}{2\sigma^2}\right),\label{eq:KRRGaussian}
\end{equation}
where $\sigma$ is the kernel width, and 
$D\left(\mathbf{M}_i,\mathbf{M}_j\right)$ is the Euclidean distance (the $L_2$ 
norm) between descriptors $\mathbf{M}_i$ and $\mathbf{M}_j$ (vectors of size 
$N_M$) defined as 
\begin{equation}
  D\left(\mathbf{M}_i,\mathbf{M}_j\right) = \sqrt{\sum_{a}^{N_M}{\left(M_{i,a} - M_{j,a}\right)^2}}\label{eq:Xdistance}
\end{equation}
This definition of distances between molecular 
structures~\cite{CMat,CMCCS,MLalgsAssessment} and a similar definition based on 
the $L_1$ norm (instead of the $L_2$ norm used 
here)~\cite{sdata2014,MLalgsAssessment,Dral_PL_2015,Dral_Delta_ML_2015,
Andrienko_ML_2015, Rupp_ML_Atoms_2015,Ramakrishnan_ML_TDDFT_2015,
Faber_2M_Crystals_2016} 
were used previously to compare molecular geometries~\cite{sdata2014} and learn 
various molecular 
properties~\cite{CMat,CMCCS,MLalgsAssessment,Dral_PL_2015,Dral_Delta_ML_2015,
Andrienko_ML_2015, 
Rupp_ML_Atoms_2015,Ramakrishnan_ML_TDDFT_2015,Faber_2M_Crystals_2016}.
\red{We have also tested the Laplacian kernel employing the $L_1$ norm,
but the PESs obtained with this kernel showed much larger errors than those obtained
with the Gaussian kernel.}

 Training the ML model involves finding the regression coefficients 
$\boldsymbol{\alpha}$, i.e. solving the minimization problem,
\begin{equation}
\min_\alpha{\sum_j^{N_{\mathrm{train}}}{\left[Y^{\mathrm{ML}}\left(\mathbf{M}_j\right) - Y^{\mathrm{ref}}\left(\mathbf{M}_j\right)\right]}} + \lambda\boldsymbol{\alpha}^T\mathbf{K}\boldsymbol{\alpha},\label{eq:KRRMinimizationProblem}
\end{equation}
which has a known analytical solution~\cite{Hastie_Book_2009}
\begin{equation}
  \boldsymbol{\alpha} = \left(\mathbf{K} + \lambda\mathbf{I}\right)^{-1} \mathbf{Y}^{\mathrm{ref}}\label{eq:KRRMinimizationProblemSolution}
\end{equation}
where $\mathbf{I}$ is the identity matrix,
\red{$\mathbf{K}$ is the kernel matrix with the elements calculated using
Eq.~\ref{eq:KRRGaussian} for all pairs of training set points}, 
$\mathbf{Y}^{\mathrm{ref}}$ is a 
vector with reference property values, and $\lambda$ is the so-called 
regularization parameter that ensures the transferability of the model to new 
nuclear configurations~\cite{CMat,CMCCS}.

 Two additional improvements were made to the approach outlined above which 
were not considered in earlier 
work:~\cite{CMat,CMCCS,MLalgsAssessment,Dral_Delta_ML_2015,Dral_PL_2015} (i) we 
sample data to ensure that ML interpolates between the training points and does 
not extrapolate beyond them (see Section~\ref{subsec:sampling}), and (ii) we 
use a nested, self-correcting ML approach. For the latter, ML is first trained 
on the reference \abinitio{} energies $\mathbf{E}^{\mathrm{ref}}$ 
($\mathbf{Y}^{\mathrm{ref}}$ in Eq.~\eqref{eq:KRRMinimizationProblemSolution}) 
and then used to make a first-layer estimate of the deformation energy 
$E^{\mathrm{layer}\,1}$ ($Y^{\mathrm{ML}}$ in Eq.~\eqref{eq:KRRest}). The next 
layer corrects errors of the energies estimated in the previous layer, and so 
on. For example, the second layer ML model is trained on $\mathbf{\Delta 
E}^{\mathrm{ref,}\,\mathrm{layer}\,1} = 
\mathbf{E}^{\mathrm{ref}}-\mathbf{E}^{\mathrm{layer}\,1}$ and is then used to 
calculate $\mathbf{\Delta E}^{\mathrm{ML,}\,\mathrm{layer}\,1}$ 
corrections for the prediction set which are summed up with layer 1 predicted 
energies to obtain layer 2 energies, i.e. $\mathbf{E}^{\mathrm{layer}\,2} = 
\mathbf{\Delta E}^{\mathrm{ML,}\,\mathrm{layer}\,1} + 
\mathbf{E}^{\mathrm{layer}\,1}$. The dependence of the performance of the ML 
model on the number of layers is discussed in 
Section~\ref{subsec:optimalmlmodel}.

 In order to determine the optimal values of the hyperparameters $\sigma$ and 
$\lambda$ for each layer, we sample 80\% of the training set points into a 
sub-training set using the same \red{ sampling procedure that was
employed for the 
training set} (see Section~\ref{subsec:sampling}). The points with deformation 
energies less than 10,000~\wn{} are taken from the remaining 20\% into the 
validation set. Using the ML model trained on the sub-training set, we search 
for values of the hyperparameters which give the lowest root-mean-square error 
(RMSE) for the validation set. This is performed using a simple logarithmic 
grid search~\cite{MLalgsAssessment,Rupp_IJQC_Tutorial_2015}. By optimizing the 
hyperparameters such as to obtain a better description below 10,000~\wn{} we 
ensure an adequate treatment of the spectroscopically most relevant part of the 
PES.\footnote{On only one occasion
(one of the ML models trained on 50\% randomly drawn grid points)
the RMSE of the training set in the first-layer became 
unreasonably high due to overfitting caused by very small value 
of the parameter $\lambda$. In this case we set $\lambda$ to $10^{-6}$ for the first layer
and re-trained the ML model.}

 The above self-correcting scheme is similar in spirit to a two-stage NN model 
that has been introduced to remedy peculiarities of 
NNs~\cite{Manzhos_nestedNN_2006,Manzhos_NN_2014}, such as the presence of 
ill-defined regions (``holes'') in NN PESs as a result of overfitting. In this 
previous work, the NN potential was first fit with as few nodes as possible to 
eliminate holes, which resulted in a large RMSE; to obtain a PES of reasonable 
quality, the NN potential was then re-fit by incorporating more nodes to 
eliminate residual errors.
 
 In our study, overfitting is primarily prevented by using the regularization 
parameter $\lambda$ and the problem of holes on the PES is addressed by our 
sampling procedure (discussed in Section~\ref{subsec:sampling}). The KRR 
approach makes fitting straightforward as there is no need to manually adjust 
any of the parameters. As a result, the RMSE of our one-layer ML model is 
already quite low, while multi-layer ML models possess still slightly smaller 
RMSEs (see Section~\ref{subsec:optimalmlmodel}). This is in contrast to 
two-stage NN models where the RMSE of the first fit is one order of magnitude 
larger than that of the second fit~\cite{Manzhos_nestedNN_2006}.

\subsection{Molecular descriptor}
\label{subsec:moldescr}

 The success of ML is largely determined by the choice of an appropriate 
molecular descriptor. Many of the molecular descriptors proposed in literature 
are functions of the internuclear distances, for example the Coulomb 
matrix\cite{CMat,MLalgsAssessment}, the Bag of Bonds\cite{Hansen_BoB_2015}, the 
atom-centered symmetry 
functions\cite{Behler_Parrinello_NN_2007,Behler_Perspective-ML-potentials_2016},
 the bispectrum of the neighbor density\cite{Bartok_GAP_2010}, the smooth 
overlap of atomic positions\cite{Bartok_SOAP_2013}, and 
others\cite{Majumder_MP_2015}. We designed and tested many descriptors for 
\ce{CH_3Cl} but overall the most accurate was a vector with ten elements 
corresponding to the ten interatomic pairs. Each element is defined as the 
corresponding internuclear distance in the near-equilibrium reference geometry 
of \ce{CH_3Cl} ($r^{\mathrm{eq}}$) divided by the current distance ($r$), e.g. 
$r^{\mathrm{eq}}_{\text{C}-\text{Cl}}/r_{\text{C}-\text{Cl}}$ for the \ce{C-Cl} 
atomic pair. This form of 
descriptor ensures that the ML model is rotationally and translationally 
invariant. 

Since the molecular descriptor also has to be atom index invariant, 
\red{we sort the} hydrogen nuclei by the sum of their internuclear repulsions with the 
four other nuclei \red{for structure-based sampling (see
Section~\ref{subsec:sampling})}.
\red{Simple sorting of hydrogen nuclei in the molecular descriptor may 
however lead to instabilities in regions where the order of hydrogen 
nuclei changes. To avoid this problem we employ
a normalized permutational invariant kernel in our ML calculations
(Section~\ref{subsec:ml}) as suggested in the
literature:\cite{Bartok_GPwater_2013,Bartok_IJQC_Tutorial_2015}
\begin{equation}
  \overline{K}\left(\mathbf{M}_i,\mathbf{M}_j\right) = 
  \frac{\sum_{\hat{P}}^{N_{perm}}{K\left(\mathbf{M}_i,\hat{P}\mathbf{M}_j\right)}}
  {\sqrt{\sum_{\hat{P}}^{N_{perm}}{K\left(\mathbf{M}_i,\hat{P}\mathbf{M}_i\right)}}
   \sqrt{\sum_{\hat{P}}^{N_{perm}}{K\left(\mathbf{M}_j,\hat{P}\mathbf{M}_j\right)}}},
   \label{eq:Kpermute}
\end{equation}
where $\hat{P}$ permutes the order of hydrogen nuclei. There are
${N_{perm}} = 3! = 6$ permutations of three hydrogen nuclei.
We found that results obtained using the normalized permutational invariant
kernel are superior to those obtained using a sorted molecular descriptor.}

\subsection{Grid points and sampling}
\label{subsec:sampling}

The target PES needs to be evaluated on a large number of pre-defined grid 
points (nuclear configurations). These grid points can be determined through 
rather inexpensive initial energy computations. In our previous work on 
\ce{CH_3Cl}, this was done by creating a sparse grid along each one-dimensional 
cut of the PES, calculating single-point energies using a reliable but 
relatively cheap \abinitio{} method (e.g. CCSD(T)-F12b/VTZ-F12), fitting a 
polynomial function to get rough energy estimates, and randomly selecting all 
remaining points using an energy-weighted Monte Carlo type sampling algorithm 
to cover the desired low-energy PES region\cite{15OwYuYa.CH3Cl}. This 
procedure, which was inexpensive computationally, produced a grid of 44,819 
geometries with energies up to $h c \cdot$~50,000~\wn{} ($h$ is the Planck 
constant and $c$ is the speed of light). In our present proof-of-principle 
study we use this grid, which was employed for the final 
CBS-35$^{\,\mathrm{HL}}$ PES of \ce{CH_3Cl}\cite{15OwYuYa.CH3Cl},
 and partition it into training and test sets using the structure-based 
sampling procedure described below; the energies for the training set are taken 
from the available \abinitio{} results\cite{15OwYuYa.CH3Cl} while those for the 
remaining grid points are predicted essentially for free using ML.
 
\begin{figure}
  \includegraphics[width=3.37in]{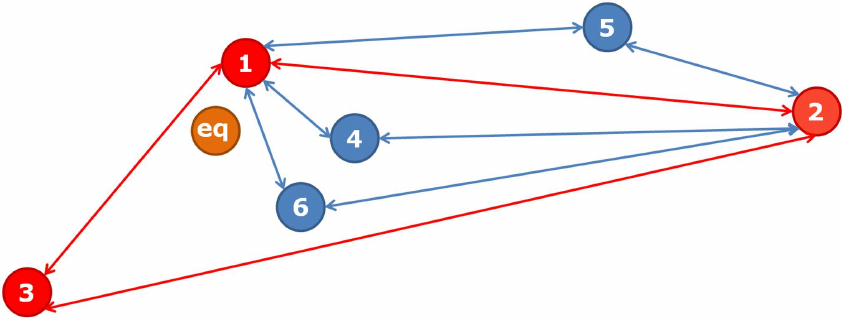}
  \caption{Illustration of sampling 3 points into the training set 
           out of 6 points in two-dimensional space.
           Point 1 is closest to the equilibrium,
           while point 2 is farthest apart from point 1,
           so they are both included into the training set.
           Point 3 has the shortest distance to point 1 of
           the current training set, which is longer than 
           any shortest distance of points 4, 5, and 6 to
           the points in the current training set (1 and 2).
           Thus, point 3 is included into the training set
           (red), while the remaining points 4--6 are left
           for the prediction set (blue).}
  \label{fgr:sampling}
\end{figure}

 In our sampling procedure we select nuclear configurations for the training 
set based on the relative distance of the molecular descriptors. This is done 
such that all remaining points used for prediction are within the boundaries of 
the training set or very close to it. The first point of the training set is 
taken closest to equilibrium. The second point of the training set is the one 
among all grid points that has the largest distance from the first point. For 
each remaining point on the grid we calculate its distance to all points in the 
training set using Eq.~\eqref{eq:Xdistance} and determine the shortest distance 
to any point in the training set. We then include the grid point that has the 
`longest' shortest distance into the training set, as illustrated in 
Figure~\ref{fgr:sampling}. This procedure is repeated until the required number 
of points are selected for the training set. The other remaining points are 
used for prediction, and by construction they lie within the training set or 
very close to 
it; at least one of their distances to the points in the training set should be 
shorter than the shortest distance between points in the training set.

\red{This sampling procedure is closely related to the farthest-point traversal
iterative procedure used to select points such that they are as distant as possible
from the previously selected points. In this respect structure-based sampling
can be also viewed as a way to obtain a training set as diverse as possible.}

 The sampling procedure outlined above selects a training set of predetermined 
size from a larger set of predefined structures. The same sampling principles 
can be applied to test whether additional structures (beyond the initially 
chosen set) should be included into the training set. If this is the case, the 
ML model needs to be re-trained (similar to an approach described in 
Ref.~\citenum{Behler_PCCP_2011}).

\subsection{Variational calculations}
\label{subsec:variational}

 In this work we use the nuclear motion program 
TROVE~\cite{TROVE2007,15YaYu.ADF} for computing vibrational energy levels. 
Since ro-vibrational calculations have previously been reported for 
CH$_3$Cl~\cite{15OwYuYa.CH3Cl,16OwYuYa.CH3Cl}, we summarize only the key 
aspects relevant for the present study.
 
 In variational calculations the PES must be represented analytically. To do 
this we introduce the coordinates,
\begin{equation}\label{eq:stretch1_ch3cl}
\xi_1=1-\exp\left(-a(r_0 - r_0^{\mathrm{eq}})\right) ,
\end{equation}
\begin{equation}\label{eq:stretch2_ch3cl}
\xi_j=1-\exp\left(-b(r_i - r_1^{\mathrm{eq}})\right){\,};\hspace{2mm}j=2,3,4{\,}, \hspace{2mm} i=j-1 ,
\end{equation}
where $a=1.65{\,}\mathrm{\AA}^{-1}$ for the C--Cl bond length $r_0$, and 
$b=1.75{\,}\mathrm{\AA}^{-1}$ for the three C--H bond lengths $r_1,r_2$ and 
$r_3$. For the angular terms,
\begin{equation}\label{eq:angular1_ch3cl}
\xi_k = (\beta_i - \beta^{\mathrm{eq}}){\,};\hspace{2mm}k=5,6,7{\,}, \hspace{2mm} i=k-4 ,
\end{equation}
\begin{equation}\label{eq:angular2_ch3cl}
\xi_8 = \frac{1}{\sqrt{6}}\left(2\tau_{23}-\tau_{13}-\tau_{12}\right) ,
\end{equation}
\begin{equation}\label{eq:angular3_ch3cl}
\xi_9 = \frac{1}{\sqrt{2}}\left(\tau_{13}-\tau_{12}\right) ,
\end{equation}
where $\beta_1$, $\beta_2$, $\beta_3$ are the 
$\angle(\mathrm{H}_i\mathrm{CCl})$ interbond angles, and $\tau_{12}$, 
$\tau_{13}$, $\tau_{23}$ are the dihedral angles between adjacent planes 
containing H$_i$CCl and H$_j$CCl. Here $r_0^{\mathrm{eq}}$, $r_1^{\mathrm{eq}}$ 
and $\beta^{\mathrm{eq}}$ are the reference equilibrium structural parameters 
of CH$_3$Cl.
 
 The potential function (maximum expansion order $i+j+k+l+m+n+p+q+r=6$) is 
given by the expression,
\begin{equation}\label{eq:pot_ch3cl}
V(\xi_{1},\xi_{2},\xi_{3},\xi_{4},\xi_{5},\xi_{6},\xi_{7},\xi_{8},\xi_{9})={\sum_{ijk\ldots}}{\,}\mathrm{f}_{ijk\ldots}V_{ijk\ldots} ,
\end{equation}
and contains the terms,
\begin{equation}
V_{ijk\ldots}=\lbrace\xi_{1}^{\,i}\xi_{2}^{\,j}\xi_{3}^{\,k}\xi_{4}^{\,l}\xi_{5}^{\,m}\xi_{6}^{\,n}\xi_{7}^{\,p}\xi_{8}^{\,q}\xi_{9}^{\,r}\rbrace^{\bm{C}_{3\mathrm{v}}\mathrm{(M)}} ,
\end{equation}
which are symmetrized combinations of different permutations of the vibrational 
coordinates $\xi_{i}$, and transform according to the $A_1$ representation of 
the $\bm{C}_{3\mathrm{v}}\mathrm{(M)}$ molecular symmetry 
group~\cite{MolSym_BuJe98}. 

\begin{figure}
  \includegraphics[width=3.37in]{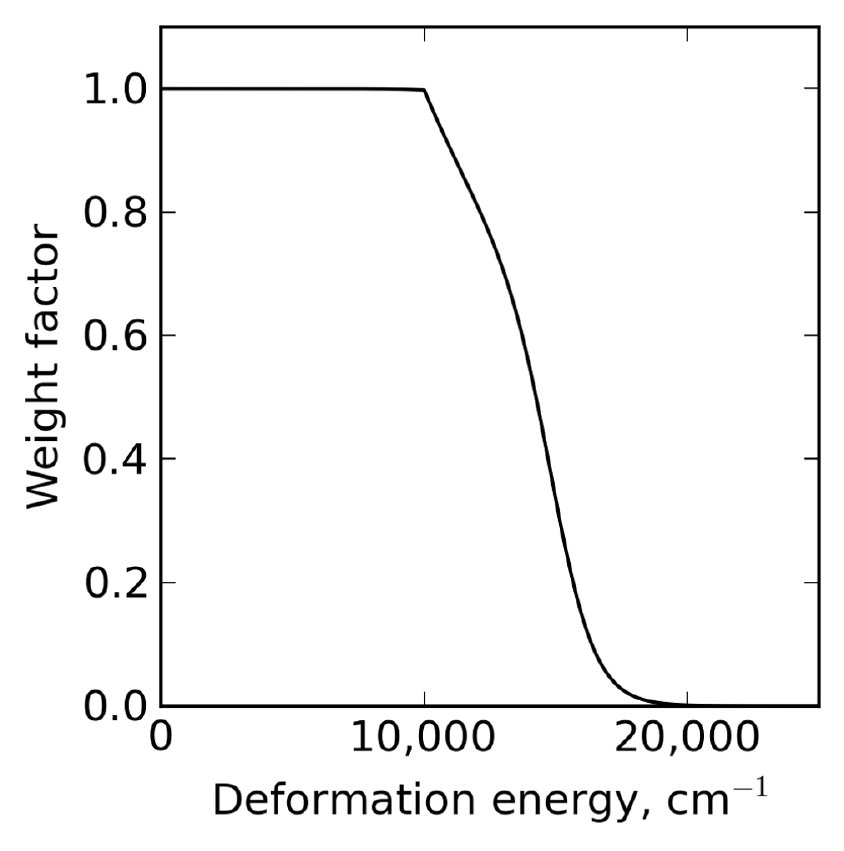}
  \caption{Decay of unitless weight factors calculated using
           Eq.~\eqref{eq:weight} with increasing deformation energy
           in \wn{}.}
  \label{fgr:weight}
\end{figure}

 The $\mathrm{f}_{ijk\ldots}$ expansion coefficients are determined through a 
least-squares fitting to the \abinitio{} and/or ML data. Weight factors 
of the form~\cite{Schwenke97}
\begin{equation}\label{eq:weight}
w\left(E_i\right)=\left(\frac{\tanh\left[-0.0006\times(E_i - 15{\,}000)\right]+1.002002002}{2.002002002}\right)\times\frac{1}{NE_i^{(w)}} ,
\end{equation}
are used in the fittings. Here $E_i^{(w)}=\max(E_i, 10{\,}000)$ and the 
normalization constant $N=0.0001$ (all values in cm$^{-1}$). Larger weights 
($w$) are assigned to lower deformation energies ($E_i$), which correspond to 
more spectroscopically important regions of the PES. As shown in 
Figure~\ref{fgr:weight}, this form ensures that structures with energies up to 
10,000~\wn{} above equilibrium have weights near unity, whilst other 
configurations are significantly downweighted with increasing energy.

To ensure a reliable comparison the CBS-35$^{\,\mathrm{HL}}$ PES expansion 
parameter set was used to fit the \abinitio{} and ML-generated datasets. 
This contained 414 parameters and included linear expansion terms. For this 
reason we fixed the values of $r_0^{\mathrm{eq}}=1.7775{\,}\mathrm{\AA}$, 
$r_1^{\mathrm{eq}}=1.0837{\,}\mathrm{\AA}$, and 
$\beta^{\mathrm{eq}}=108.445^{\circ}$, to be the same as those used for the 
CBS-35$^{\,\mathrm{HL}}$ PES. Each fit employed Watson's robust fitting 
scheme~\cite{Watson03}, which reduces the weights of outliers and lessens their 
influence in determining the final set of parameters. 
 
\section{Results and discussion}
\label{sec:results}

  In this section we investigate the accuracy of the ML model\red{s} trained on 50\%, 
25\%, and 10\% of the available 44,819 deformed structures and their 
\abinitio{} energies used to construct the CBS-35$^{\,\mathrm{HL}}$ 
PES~\cite{15OwYuYa.CH3Cl}.

 To evaluate the accuracy of predicted energies ($E^{\text{est}}$) we employ 
standard error measures as well as the weighted root-mean-square error (wRMSE):
 \begin{equation}\label{eq:wRMSE}
\text{wRMSE}=\sqrt{\frac{1}{N}\sum_i^N{\left(E_i^{\text{est}} - E_i^{\text{ref}} \right)^2 w\left(E_i^{\text{ref}} \right) }}.
\end{equation}
 Weights $w\left(E_i^{\text{ref}}\right)$ are calculated using 
Eq.~\ref{eq:weight}.

\subsection{Optimal machine learning model}
\label{subsec:optimalmlmodel}

 \red{We first examine how the accuracy of ML calculated deformation energies for 
the prediction set depends on the number of ML layers 
and the sampling procedure. The results in 
Table~\ref{tbl:ml-sampling-layers} indicate that the errors are 
significantly lowered by adding the second layer for all models except for
the one trained on 10\% points selected with structure-based sampling.
Including a third layer significantly reduces the errors in a few cases,
while a fourth layer does not yield any 
noticeable improvements. We therefore expect that adding further layers will 
make no difference.}

\red{\setlength\LTleft{0pt}
\setlength\LTright{0pt}
\LTcapwidth=\textwidth
\begin{longtable*}[ht]{@{\extracolsep{\fill}}r c c c c c c}
\caption{\label{tbl:ml-sampling-layers} wRMSEs in deformation energies 
         predicted by ML models
         trained on 50\%, 25\%, and 10\% of the reference data
         for the remaining 50\%, 75\%, and 90\% of the grid points.
         wRMSEs are weighted according to Eq.~\eqref{eq:weight} and
         are given in \wn{}.
         The number of training set structures with an \abinitio{} 
         deformation energy
         below 1,000~\wn{} ($N_{<1{,}000}$) and below 10,000~\wn{} 
         ($N_{<10{,}000}$)
         is also given for comparison.}\\ \hline\hline
    \multicolumn{1}{c}{$N_{\mathrm{train}}$ } & $N_{<1{,}000}$ $^a$ & $N_{<10{,}000}$ $^b$ & \multicolumn{4}{c}{Number of layers} \\ \cline{4-7} 
                                         &                &                 & 1 & 2 & 3 & 4 \\ \hline
\endfirsthead
\caption{(\textit{Continued})}\\ \hline 
    \multicolumn{1}{c}{$N_{\mathrm{train}}$ } & $N_{<1{,}000}$ $^a$ & $N_{<10{,}000}$ $^b$ & \multicolumn{4}{c}{Number of layers} \\ \cline{4-7} 
                                         &                &                 & 1 & 2 & 3 & 4 \\ \hline
\endhead
    \hline
    \multicolumn{7}{c}{Structure-based sampling from unsliced data} \\
    \hline
     22,409 (50\%) &    22  &  3,985  &     6.57  &     0.37  &     0.37  &     0.37  \\
     11,204 (25\%) &     1  &  1,033  &     3.21  &     1.87  &     1.60  &     1.60  \\ 
      4,481 (10\%) &     1  &    195  &      5.00  &     4.99  &     4.83  &     4.83  \\
    \hline
    \multicolumn{7}{c}{Structure-based sampling from data sliced into three regions} \\
    \hline
     22,408 (50\%) &   131  &  7,215  &     2.31  &     0.62  &     0.62  &     0.62  \\
     11,203 (25\%) &    19  &  3,348  &     2.90  &     2.59  &     2.58  &     2.58  \\ 
      4,480 (10\%) &     1  &  1,191  &     4.42  &     3.63  &     3.63  &     3.63  \\
    \hline
    \multicolumn{7}{c}{Random sampling$^c$} \\
    \hline
     22,409 (50\%) & $573\pm12$ & $7{,}971\pm53$ &    $6.07\pm2.96$ & $4.13\pm0.87$ & $4.13\pm0.87$ & $4.13\pm0.87$ \\
     11,204 (25\%) & $288\pm13$ & $4{,}002\pm36$ &    $8.41\pm8.23$ & $4.76\pm0.90$ & $4.75\pm0.88$ & $4.75\pm0.88$ \\       
      4,481 (10\%) & $115\pm12$ & $1{,}599\pm35$ &  $14.73\pm12.63$ & $8.76\pm1.67$ & $8.73\pm1.67$ & $8.73\pm1.67$ \\      
    \hline\hline
  \multicolumn{7}{c}{$^a$~$N_{< 1{,}000} =  1{,}145$ for the entire grid of 44,819 points.}\\
  \multicolumn{7}{c}{$^b$~$N_{<10{,}000} = 15{,}935$ for the entire grid of 44,819 points.}\\
  \multicolumn{7}{c}{$^c$~Standard deviations were calculated for 16, 20, and 30 various randomly drawn}\\
  \multicolumn{7}{c}{training sets for ML trained on 50\%, 25\%, and 10\% grid points, respectively.}\\
\end{longtable*}}

 For practical purposes the three-layer model appears to be sufficient. 
However, in the following we have applied the four-layer model because the 
computational cost of including additional layers is rather low \red{and we have not observed any significant accumulation of numerical noise}. It takes 
around 4 hours to optimize the ML hyperparameters using a fairly inefficient 
optimization algorithm, around a minute to train and only a couple of seconds 
to make predictions with the 50\%-ML model on 20 cores of an Intel(R) Xeon(R) 
CPU E5-2687W v3 @ 3.10GHz processor.

 As shown in Table~\ref{tbl:ml-sampling-layers}, random sampling is 
clearly inferior to structure-based sampling; all four-layer ML models trained 
on randomly drawn points have wRMSEs significantly higher than those of the 
four-layer, structure-based sampling ML models. Interestingly, self-correction 
works well even for random sampling: it reduces the wRMSEs by \red{32--44}\% and the 
standard deviations by \red{70--90}\%. Despite this the remaining error of 
\red{$4.13\pm0.87$}~\wn{} for the randomly sampled, four-layer ML models trained on 
50\% of grid points is still much higher than the wRMSE of \red{0.37~\wn{} for the 
four-layer, structure-based sampling ML model trained on the same number of grid points}. Standard deviations for random 
sampling are also relatively high and increase from \red{0.87} to \red{1.67}~\wn{} when 
going from the 50\% to 10\% training sets. There is no such problem with 
structure-based sampling which provides a unique training set --- an important 
practical advantage for high-accuracy applications. As for the computational
cost of sampling, it takes ca. 9 hours to sample 50\% from 44,819 data points on
12 Intel(R) Xeon(R) CPU X5680 @ 3.33GHz processors.

\begin{figure}
  \includegraphics[width=3.37in]{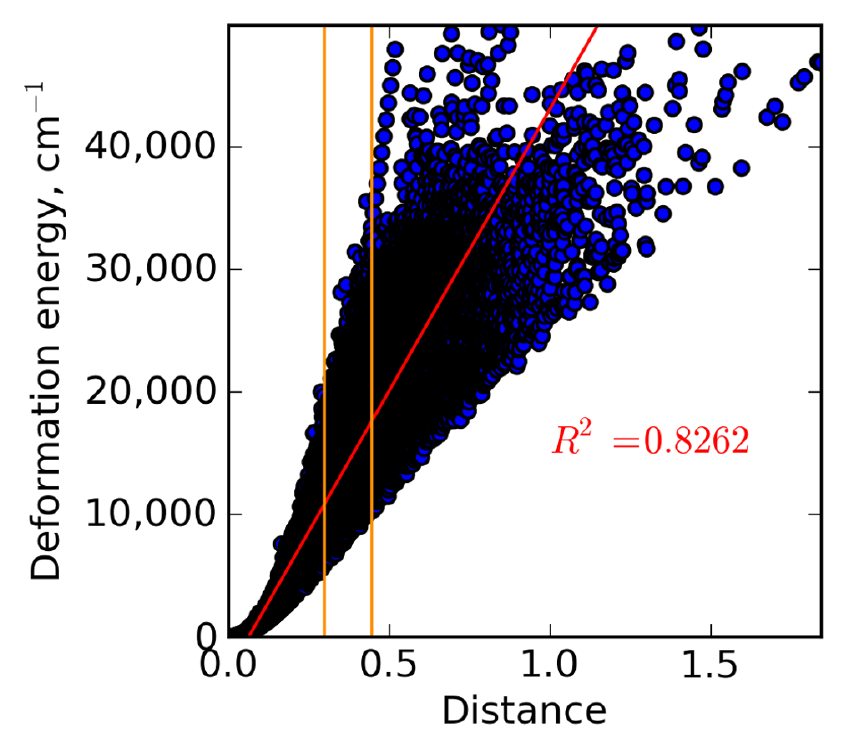}
  \caption{Correlation between \abinitio{} deformation energies
           in \wn{} and unitless distances to the near-equilibrium
           structure calculated using Eq.~\eqref{eq:Xdistance}.
           A linear trend line is shown in red with its $R^2$ value
           (0.83).
           Orange vertical lines slice the data into three regions
           with equal numbers of data points in the training set.
           Each data point is represented by a blue dot with a
           black edge, hence the most densely populated areas are black.}
  \label{fgr:slicing}
\end{figure}

\red{As the training set becomes relatively small, structure-based sampling
may lead to an under-representation of the training points in the low-energy
region, e.g. 10\% training points drawn from the entire grid contain only
195 structures with deformation energies below 10,000~\wn{} 
(compare the number of configurations with deformation energies below
1,000 and 10,000~\wn{}, columns $N_{<1{,}000}$ and $N_{<10{,}000}$ in
Table~\ref{tbl:ml-sampling-layers}).
Sorting the geometries by their distance to the
equilibrium structure (which correlates strongly with the deformation energies, Figure~\ref{fgr:slicing}),
followed by slicing the data into several regions and sampling
points from each of these regions leads to a more energy-balanced training set.
Looking at Figure~\ref{fgr:slicing}, one can argue that splitting
the set into three regions with equal number of structures should be close to optimal.
The first region includes the most important 
structures with deformation energies below 10,000~\wn{} and a significant 
portion of structures with energies between 10,000 and 20,000~\wn{}. The second 
region mainly includes configurations with energies between 10,000 and 
20,000~\wn{} but also a considerable number of geometries above and below this 
region. The third slice includes all remaining high-energy structures. 
Structure-based sampling from each of the above regions leads to $N_{<10{,}000}$ close
to those expected from random sampling,
e.g. 10\% training points drawn from the sliced grid contain 
1,191 structures with deformation energies below 10,000~\wn{} (Table~\ref{tbl:ml-sampling-layers}).
As a result, wRMSE of the ML model trained on the latter training set (3.63~\wn{}) is lower than
wRMSE of the ML model trained on 10\% points drawn from unsliced data (4.83~\wn{}).
However, such slicing does not generate training sets
that are as diverse as possible, and therefore the errors of the ML models
become higher for the sliced grids as the training set increases in size
(training sets with 25\% and 50\% grid points, Table~\ref{tbl:ml-sampling-layers}).
Thus we recommend slicing only for very small training sets where
low-energy structures are under-represented.}

\red{In the following we will discuss four-layer ML 
models trained on 50\%, 25\%, and 10\% points drawn using structure-based sampling from the available unsliced 44,819 grid
points. We refer to these ML models as 50\%-ML, 25\%-ML, and 10\%-ML, respectively. We also compare with one
of the randomly sampled, four-layer ML models referred to as r50\%-ML which has 
been chosen at random from the ML models trained on 50\% of grid points. In addition,
we compare with the four-layer ML model trained on 10\% points drawn using structure-based sampling
from each region of the dataset sliced into three regions. This model is referred to as s10\%-ML in the following.}
 
 A more detailed analysis of the ML model errors, listed in 
Table~\ref{tbl:ml-statistics}, reveals that r50\%-ML has the largest outliers 
with a wRMSE of \red{4.14}~\wn, which is \red{more than} twice as large as that of 25\%-ML 
(wRMSE of \red{1.60}~\wn) for their respective prediction sets. The non-weighted RMSE 
of r50\%-ML (\red{167.19}~\wn) is \red{more than four times higher than} the non-weighted RMSE of 10\%-ML 
(\red{39.63}~\wn). Moreover, the RMSE of r50\%-ML for energies below 1,000~\red{\wn{} is practically the same as the respective RMSE of 50\%-ML and for energies below 
10,000~\wn{} is higher than the respective RMSE of 50\%-ML}, despite the fact 
that \red{many} more points from these regions are included into the training set (compare 
$N_{<1{,}000}$ and $N_{<10{,}000}$ for these two models in 
Table~\ref{tbl:ml-sampling-layers}). These observations provide strong 
evidence for the superiority of structure-based sampling. \red{s10\%-ML has RMSEs
for energies below 1,000 and 10,000~\wn{} close to those of 25\%-ML, while the respective
RMSEs of 10\%-ML are much higher. On the other hand, 10\%-ML
has a much lower non-weighted RMSE, MAE and MSE than s10\%-ML. Thus, slicing clearly improves the description
of the low-energy region at the cost of other regions. Apparently, for sparse training data, the benefits
of a better description of the low-energy region achieved by slicing outweigh the disadvantage of an overall worse description of the PES, which
is exemplified by the lower wRMSE of s10\%-ML compared to the wRMSE of 10\%-ML.}

\red{
\setlength\LTleft{0pt}
\setlength\LTright{0pt}
\LTcapwidth=\textwidth
\begin{longtable*}[ht]{@{\extracolsep{\fill}}l r r r r r}
\caption{\label{tbl:ml-statistics} Number of grid points in the training 
($N_{\mathrm{train}}$) and
           prediction ($N_{\mathrm{predict}}$) sets, largest positive (LPO)
           and negative (LNO) outliers, mean signed errors (MSEs),
           mean absolute errors (MAEs), RMSEs for entire sets (all),
           for structures with reference deformation energies below
           1,000 and 10,000~\wn{}, and wRMSEs in \wn{} for the
           training and prediction sets of the 50\%-ML, r50\%-ML,
           25\%-ML, 10\%-ML, and s10\%-ML models and the entire grid of 44,819 points.}\\ \hline\hline
                     & 50\%-ML & r50\%-ML &  25\%-ML &  10\%-ML & s10\%-ML \\ \hline
\endfirsthead
\caption{(\textit{Continued})}\\ \hline 
                     & 50\%-ML & r50\%-ML &  25\%-ML &  10\%-ML & s10\%-ML \\ \hline
\endhead
    \hline
    \multicolumn{5}{c}{Training set} \\
    \hline
    $N_{\mathrm{train}}$       &   22,409 &    22,409 &    11,204 &     4,481 & 4,480 \\
    LPO              &    0.00 &     4.18 &     0.00 &     0.00 &     0.00 \\
    LNO              &    0.00 &   $-4.14$&     0.00 &     0.00 &     0.00 \\
    MSE              &    0.00 &     0.00 &     0.00 &     0.00 &     0.00 \\
    MAE              &    0.00 &     0.03 &     0.00 &     0.00 &     0.00 \\
    RMSE (all)       &    0.00 &     0.11 &     0.00 &     0.00 &     0.00 \\
    RMSE ($<$10,000) &    0.00 &     0.03 &     0.00 &     0.00 &     0.00 \\
    RMSE ($<$1,000)  &    0.00 &     0.07 &     0.00 &     0.00 &     0.00 \\
    wRMSE            &    0.00 &     0.03 &     0.00 &     0.00 &     0.00 \\
    \hline
    \multicolumn{5}{c}{Prediction set} \\
    \hline
    $N_{\mathrm{predict}}$    &   22,410 &    22,410 &    33,615 &    40,338 &     40,339 \\
    LPO              & 319.75 & 2,015.44 &  1,035.63 &  1,617.61 &  1,481.69 \\
    LNO              &$-476.20$&$-6{,}919.28$&$-1{,}060.38$&$-2{,}859.37$&$-2{,}190.33$ \\
    MSE              &    0.02 &  $-11.62$&     0.18 &     0.19 &     1.85 \\
    MAE              &    0.82 &    25.12 &     3.47 &    11.27 &    19.54 \\
    RMSE (all)       &    6.23 &   167.19 &    16.12 &    39.63 &    61.96 \\
    RMSE ($<$10,000) &    0.20 &     1.38 &     1.12 &     4.26 &     1.19 \\
    RMSE ($<$1,000)  &    0.08 &     0.07 &     0.16 &     1.40 &     0.25 \\
    wRMSE            &    0.37 &     4.14 &     1.60 &     4.83 &     3.63 \\
    \hline
    \multicolumn{5}{c}{Entire grid} \\
    \hline
    $N$    &   \multicolumn{4}{c}{44,819} \\
    LPO              &  319.75 & 2,015.44 & 1,035.63 &  1,617.61 & 1,481.69 \\
    LNO              &$-476.20$&$-6{,}919.28$&$-1{,}060.38$&$-2{,}859.37$&$-2{,}190.33$ \\
    MSE              &    0.01 &   $-5.81$&     0.14 &     0.17 &      1.66 \\
    MAE              &    0.41 &    12.57 &     2.60 &    10.14 &     17.58 \\
    RMSE (all)       &    4.41 &   118.22 &    13.96 &    37.60 &     58.79 \\
    RMSE ($<$10,000) &    0.17 &     0.98 &     1.09 &     4.24 &     1.14 \\
    RMSE ($<$1,000)  &    0.08 &     0.07 &     0.16 &     1.40 &     0.25 \\
    wRMSE            &    0.26 &     2.93 &     1.39 &     4.59 &     3.44 \\
    \hline\hline
\end{longtable*}}

\begin{figure}
  \includegraphics[width=6.69in]{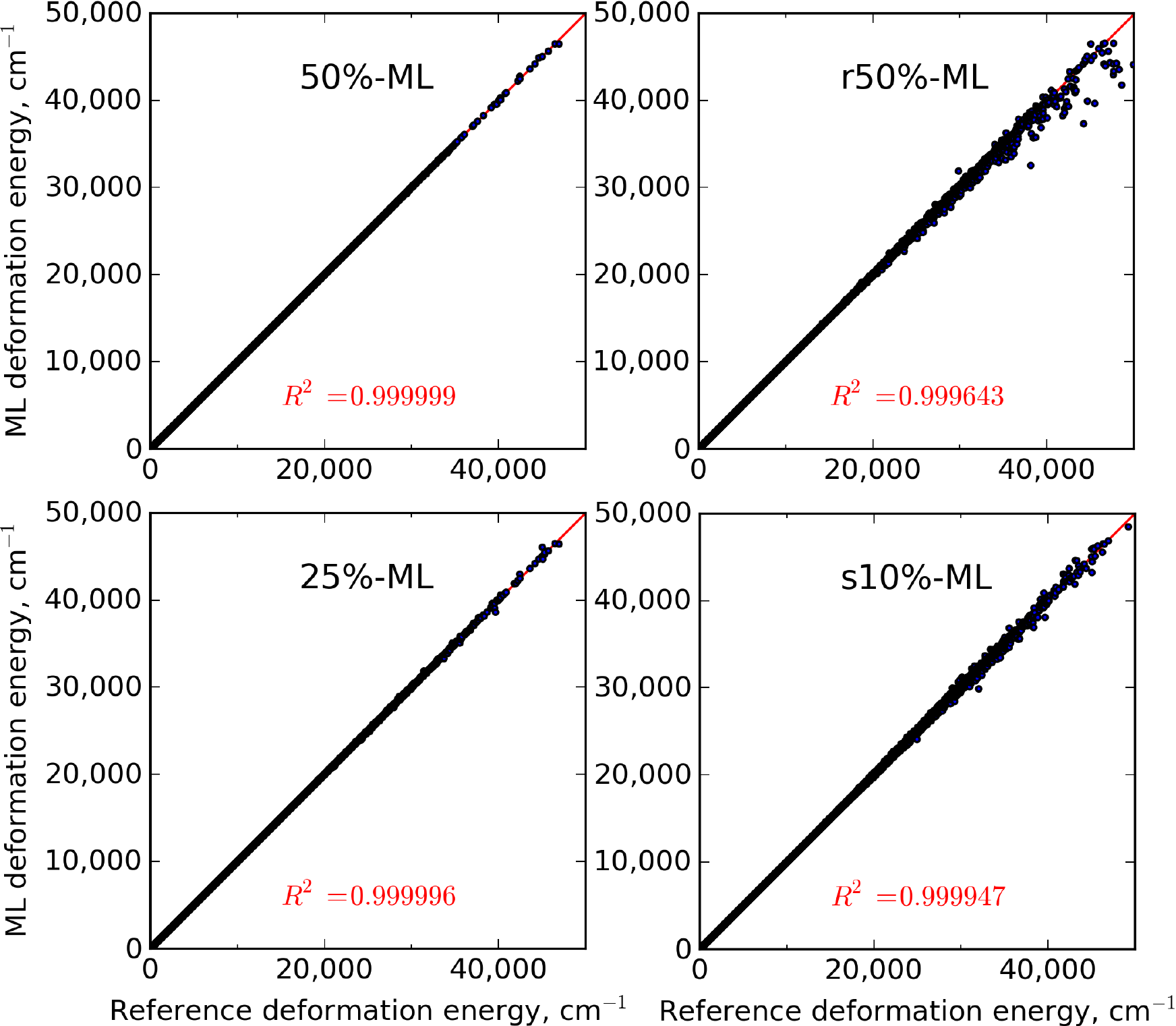}
  \caption{Correlation between reference \abinitio{}
           deformation energies and deformation energies predicted by
           50\%-ML, r50\%-ML, 25\%-ML, and \red{s}10\%-ML for their
           respective prediction sets. Linear trend lines are shown
           in red with their $R^2$ value. Each data point is
           represented by a blue dot with a black edge, hence the most
           densely populated areas are black.}
  \label{fgr:MLcorrplots}
\end{figure}

\red{From Figure~\ref{fgr:MLcorrplots} we see that deformation energies predicted 
by 50\%-ML, r50\%-ML, 25\%-ML, and s10\%-ML models correlate nicely 
with the reference \abinitio{} energies; the $R^2$ value is always 
larger than 0.999. Deformation energies predicted by 10\%-ML model (not shown in Figure~\ref{fgr:MLcorrplots})
correlate better ($R^2=0.999977$) with the reference energies than the energies predicted by the s10\%-ML model (consistent with the above conclusions).
Clearly, 50\%-ML is superior to all other models and 
has the best correlation with far fewer outliers and smaller residual errors. 
This is particularly relevant for high-accuracy work as we will see in 
Section~\ref{subsec:vibrations}.}

\begin{figure}
  \includegraphics[width=3.37in]{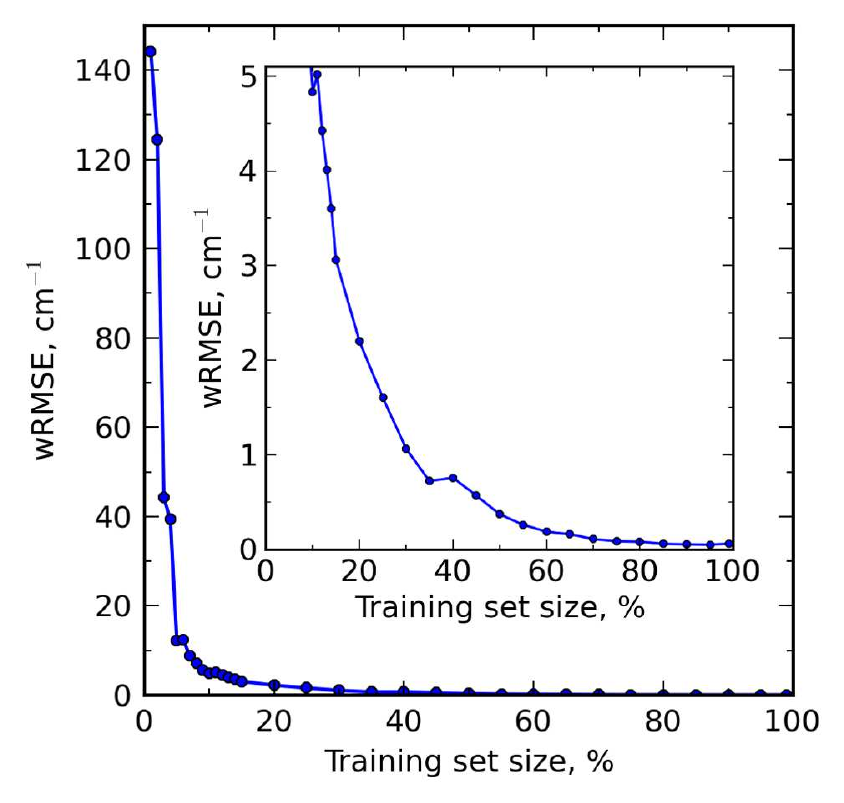}
  \caption{Dependence of wRMSE in the prediction set (in~\wn{}) of the 
four-layer, structure-based sampling ML models
           as a function of the training set size (in \%). \red{In all cases 
sampling was done from unsliced data.}
           The plot starts with a 1\% training set size and ends at 99\%. The 
plot in the inset starts with a 10\% training set size.}
  \label{fgr:MLlearningCurve}
\end{figure}

 As for the effect of training set size, illustrated in 
Figure~\ref{fgr:MLlearningCurve}, wRMSEs in the prediction set drop from 
\red{144}~\wn{} for the 1\%-ML model to \red{0.05--0.06~\wn{} for the 85--99\%-ML models}. 
Interestingly, the 1\%-ML model trained on only \red{448} grid points may still be 
regarded as a chemically meaningful representation of the PES since its 
non-weighted RMSE is only \red{0.77}~kcal/mol (\red{271}~\wn{}). The error drops very 
quickly to \red{12.29}~\wn{} for the 5\%-ML model and is below 1.00~\wn{} for \red{35}\% 
and above, finally becoming smaller than 0.5~\wn{} for training sets \red{with 50\% or more} of all configurations.
 
 Regarding the \red{four} structure-based sampling ML models that were tested 
extensively, wRMSEs grow significantly from 50\%-ML (\red{0.37}~\wn{}) to 25\%-ML 
(\red{1.60}~\wn{}) \red{to s10\%-ML (3.63~\wn{})} to 10\%-ML (\red{4.83}~\wn{}) .
\red{We investigate further the effect
of the training set size for high-resolution spectroscopy 
applications in Section~\ref{subsec:vibrations},
where we report vibrational energies using PESs based on the 50\%-ML, r50\%-ML, 
25\%-ML, 10\%-ML, and s10\%-ML models.}
 
 The analytic representation employed for the CBS-35$^{\,\mathrm{HL}}$ 
(discussed in Sections~\ref{subsec:variational} and \ref{subsec:vibrations}) 
was fitted with a wRMSE of 3.00~\wn{} for the entire grid (see 
Section~\ref{subsec:vibrations}). For 50\%-ML the training set, prediction set, 
and entire grid of 44,819 points were reproduced with wRMSEs of \red{0.00}, \red{0.37} and 
\red{0.26}~\wn{}, respectively (Table~\ref{tbl:ml-statistics}). Our ML approach, 
which has a defined analytic form, could therefore provide a more accurate 
description of the PES based on fewer training points and could possibly be 
employed directly in variational calculations. However, this is beyond the 
scope of this work and application of our ML technique in variational 
calculations will be the focus of future research.

\red{As indicated by one reviewer, further fine-tuning of the ML models is possible
by using the anisotropic Gaussian kernel with multiple kernel widths instead of
a single $\sigma$ parameter (as for instance in
Ref.~\citenum{Bartok_GPwater_2013}). In test calculations we have found that such ML models may reduce errors
somewhat (e.g. by 10\% for 10\%-ML), but there is a substantial increase in the complexity of
the ML model and the computational cost
for large training sets  (due to parameter optimization), and hence we decided to use one single kernel width
parameter in this study.}

\subsection{Vibrational energy levels}
\label{subsec:vibrations}

 For the 50\%-ML, r50\%-ML, 25\%-ML, \red{10\%-ML, and s10\%-ML} PESs, the 
$\mathrm{f}_{ijk\ldots}$ expansion coefficients of the potential function given 
in Eq.~\eqref{eq:pot_ch3cl} were determined in a least-squares fitting to the 
44,819 grid points. The results of the fittings are listed in 
Table~\ref{tab:pes_fits}. In addition, PESs were determined for the \red{five}, 
associated training sets and the results are also included in 
Table~\ref{tbl:ml-statistics}). We see that the fits of the 
CBS-35$^{\,\mathrm{HL}}$ PES, the ML model PESs, and the training set PESs are 
of a similar accuracy, with the exception of the 10\%-ML fits which exhibit 
significantly larger errors. For the other fits, the 
RMSEs below 10,000~\wn{} range between \red{1.17 and 2.08}~\wn{}, with wRMSE values 
between \red{3.01 and 3.59}~\wn{} for the ML-based PESs and up to 3.60~\wn{} for the 
training set PESs. The mean errors are particularly low for the \red{50\%-ML PES,
its training set PES, and 25\%-ML PES (0.16, 0.24, and 0.02~\wn{}, respectively).

\setlength\LTleft{0pt}
\setlength\LTright{0pt}
\LTcapwidth=\textwidth
\begin{longtable*}[ht]{@{\extracolsep{\fill}}l r r r r r r}
\caption{\label{tab:pes_fits} Number of expansion parameters 
$\mathrm{f}_{ijk\ldots}$ and fitting wRMSEs$^a$ (in \wn) for the fits of the 
CBS-35$^{\,\mathrm{HL}}$ PES, ML model PESs, and training set PESs.
           Largest positive (LPO)
           and negative (LNO) outliers, mean signed errors (MSEs),
           mean absolute errors (MAEs), RMSEs for entire sets (all),
           for structures with reference deformation energies below
           1,000 and 10,000~\wn{}, and wRMSEs in \wn{} of the fitted functions 
in respect to the entire grid of the CBS-35$^{\,\mathrm{HL}}$ energies.}\\ \hline\hline
                     & CBS-35$^{\,\mathrm{HL}}$ & 50\%-ML & r50\%-ML &  25\%-ML &  10\%-ML & s10\%-ML \\ \hline
\endfirsthead
\caption{(\textit{Continued})}\\ \hline 
                     & CBS-35$^{\,\mathrm{HL}}$ & 50\%-ML & r50\%-ML &  25\%-ML &  10\%-ML & s10\%-ML \\ \hline
\endhead
    \hline
    \multicolumn{7}{c}{ML model PESs} \\
    \hline
    No. of parameters&     414 &      414 &      414 &      412 &      402 &      409 \\
    fitting wRMSE    &    0.82 &     0.83 &     0.89 &     0.99 &     1.39 &     1.13 \\
    LPO              &2,717.33 & 2,718.84 & 2,546.10 & 2,700.54 & 2,506.24 & 2,868.59 \\
    LNO              &$-6{,}039.88$&$-6{,}039.02$&$-6{,}033.13$&$-6{,}023.04$&$-5{,}826.77$ &$-5{,}716.54$ \\
    MSE              &    0.20 &     0.16 &     0.56 &      0.02&     0.81 &     7.38 \\
    MAE              &   20.82 &    20.83 &    21.01 &    21.83 &    31.73 &    24.63 \\
    RMSE (all)       &  102.24 &   102.28 &   102.13 &   102.69 &   113.78 &   101.39 \\
    RMSE ($<$10,000) &    1.18 &     1.18 &     1.22 &     1.22 &     3.66 &     1.37 \\
    RMSE ($<$1,000)  &    0.33 &     0.33 &     0.33 &     0.34 &     1.08 &     0.40 \\
    wRMSE            &    3.00 &     3.01 &     3.09 &     3.19 &     5.40 &     3.59 \\
    \hline
    \multicolumn{7}{c}{Training set PESs} \\
    \hline
    No. of parameters&         &     414  &   414    &      414 &      410 &      411 \\
    fitting wRMSE    &         &     0.98 &   0.82   &     0.94 &     0.50 &     0.99 \\
    LPO              &         & 2,548.77 & 2,813.94 & 2,368.88 & 1,963.23 & 2,431.71 \\
    LNO              &         &$-6{,}128.55$&$-6{,}033.18$&$-6{,}185.91$&$-6{,}267.97$ &$-6{,}214.42$ \\
    MSE              &         &     0.24 &     0.75 &     1.50 &   $-4.04$&     1.35\\
    MAE              &         &    21.03 &    21.00 &    21.95 &    24.11 &    23.94 \\
    RMSE (all)       &         &   103.82 &   102.55 &   105.37 &   105.88 &   109.00 \\
    RMSE ($<$10,000) &         &     1.17 &     1.18 &     1.30 &     2.08 &     1.18 \\
    RMSE ($<$1,000)  &         &     0.35 &     0.33 &     0.33 &     0.62 &     0.32 \\
    wRMSE            &         &     3.01 &     3.02 &     3.15 &     3.60 &     3.58 \\
    \hline\hline
  \multicolumn{7}{c}{$^a$ Fitting wRMSEs are relative to the PES being fitted and not to the CBS-35$^{\,\mathrm{HL}}$ data.}\\
  \multicolumn{7}{c}{Note also that the weights differ slightly from Eq.~\ref{eq:weight} because Watson's}\\
  \multicolumn{7}{c}{robust fitting scheme~\cite{Watson03} was employed (see Section~\ref{subsec:variational}).}\\
 \end{longtable*}}

 In TROVE calculations the Hamiltonian was represented as a power series 
expansion around the equilibrium geometry in terms of nine vibrational 
coordinates and was constructed numerically using an automatic differentiation 
method~\cite{15YaYu.ADF}. The coordinates used were identical to those given in 
Eqs.~\eqref{eq:stretch1_ch3cl} to \eqref{eq:angular3_ch3cl}, except for the 
kinetic energy operator where linear expansion terms, e.g. 
$(r-r^{\mathrm{eq}})$, replace the Morse oscillator functions for the 
stretching modes.
 The kinetic and potential energy operators were truncated at 
6\textsuperscript{th} and 8\textsuperscript{th} order, respectively, and atomic 
mass values were employed throughout. Calculations were carried out using a 
medium-sized vibrational basis set with a polyad truncation number of 
$P_{\mathrm{max}}=10$ (see Refs.~\citenum{15OwYuYa.CH3Cl,16OwYuYa.CH3Cl} for 
further details). The basis set was constructed using a multi-step contraction 
scheme and contained 16,829 vibrational basis functions.

\begin{figure}
  \includegraphics[width=6.69in]{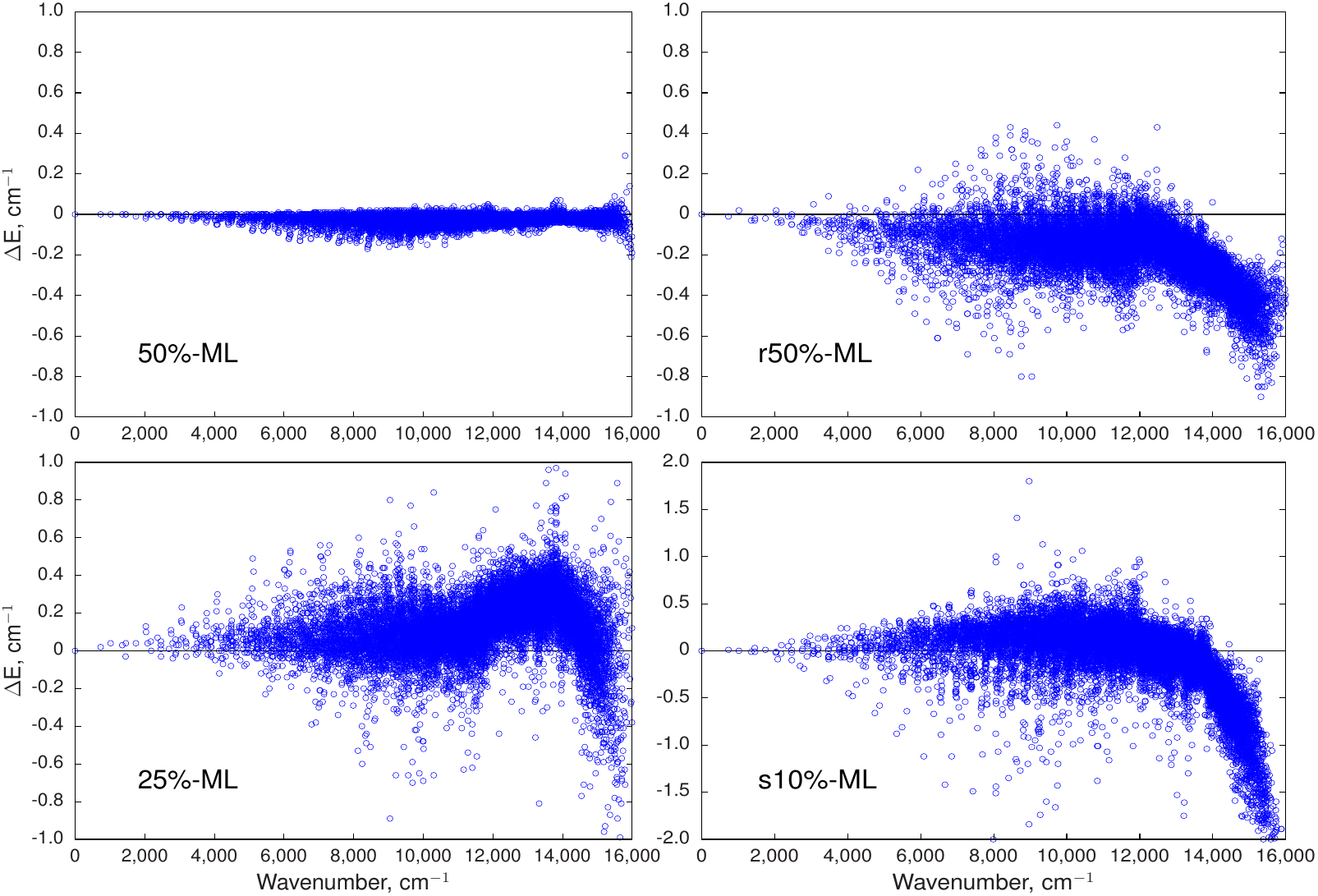}
  \caption{Residual errors ($\Delta 
E=E_{\mathrm{CBS-35^{\,HL}}}-E_{\mathrm{ML}}$) of computed vibrational energy 
levels using the 50\%-ML, r50\%-ML, 25\%-ML, and \red{s}10\%-ML PESs, with respect to 
the CBS-35$^{\,\mathrm{HL}}$ PES values. Note the different scale for 10\%-ML.}
  \label{fgr:j0_energies}
\end{figure}

 In Figure~\ref{fgr:j0_energies} we plot residual errors, $\Delta 
E=E_{\mathrm{CBS-35^{\,HL}}}-E_{\mathrm{ML}}$, of computed vibrational energy 
levels using the \red{50\%-ML, r50\%-ML, 25\%-ML, and s10\%-ML} model PESs with respect to the CBS-35$^{\,\mathrm{HL}}$ PES 
values. The RMSE and mean-absolute-deviation (MAD) for energies up to 5,000 and 
10,000~\wn{} are also listed for each model in Table~\ref{tab:j0_rms}. 
Comparing 50\%-ML with r50\%-ML, it is clear that structure-based sampling 
produces results that are far more reliable than random sampling. The residual 
errors are consistently smaller and more uniform for the energy range 
considered. The \red{25\%-ML and s10\%-ML models still perform} reasonably well but errors 
steadily increase with energy. \red{The 10\%-ML model (not shown in Figure~\ref{fgr:j0_energies}) has 
deteriorated and no longer gives accurate predictions (errors are much higher than 1~\wn{}, Table~\ref{tab:j0_rms})}.

\begin{table}[!ht]
  \footnotesize
\tabcolsep=0.25cm
\caption{\label{tab:j0_rms} Root-mean-square error (RMSE) and 
mean-absolute-deviation (MAD) of computed vibrational energy levels for the ML 
model PESs and training set PESs up to 5,000 and 10,000~\wn{} \red{(166 and 3,606 levels, respectively)}, with respect to 
the original CBS-35$^{\,\mathrm{HL}}$ PES values.}
\begin{center}
\red{\begin{tabular}{l c c c c c}
\hline\hline
& 50\%-ML & r50\%-ML & 25\%-ML & 10\%-ML & s10\%-ML\\
    \hline
    \multicolumn{6}{c}{ML model PESs} \\
    \hline
RMSE ($<$5,000\,\wn{})  & 0.02 & 0.10 & 0.09 & 1.61 & 0.14 \\
MAD ($<$5,000\,\wn{})   & 0.01 & 0.08 & 0.07 & 1.29 & 0.10 \\
RMSE ($<$10,000\,\wn{}) & 0.04 & 0.18 & 0.16 & 1.75 & 0.28 \\
MAD ($<$10,000\,\wn{})  & 0.03 & 0.15 & 0.12 & 1.44 & 0.21 \\
    \hline
    \multicolumn{6}{c}{Training set PESs} \\
    \hline
RMSE ( $<$5,000\,\wn{}) & 0.06 & 0.08 & 0.12 & 0.30 & 0.12 \\
MAD  ( $<$5,000\,\wn{}) & 0.05 & 0.06 & 0.10 & 0.25 & 0.10 \\
RMSE ($<$10,000\,\wn{}) & 0.12 & 0.14 & 0.19 & 0.74 & 0.32 \\
MAD  ($<$10,000\,\wn{}) & 0.11 & 0.09 & 0.14 & 0.61 & 0.24 \\
\hline\hline
\end{tabular}}
\end{center}
\end{table}

 TROVE assigns quantum numbers to the eigenvalues by analyzing the contribution 
from the basis functions. This is how we match energy levels computed with 
different ML model PESs. However, given the approximate nature of the labeling 
scheme these assignments can occasionally differ between surfaces. This tends 
to happen mostly at higher energies (above 10,000~\wn{}) but does not 
necessarily mean that the energy levels are mismatched --- they have simply 
been labeled differently. \red{This occurs for 2\% of the computed values in 
the case of 50\%-ML, 9\% for r50\%-ML, 12\% for 25\%-ML, and 18\% for s10\%-ML. For 10\%-ML this 
percentage rises dramatically to 46\% providing further evidence that this ML 
model is no longer reliable for high-accuracy applications.}

 We also computed vibrational energies using the training set PESs (also listed 
in Table~\ref{tab:j0_rms}). The errors are reasonably small and structure-based 
sampling again performs better than random sampling. \red{For 50\%-ML, 25\%-ML, and s10\%-ML
the predicted spectra are more reliable than the results of the respective 
training set PESs (constructed from 50\%, 25\%, and 10\% of the entire
CBS-35$^{\,\mathrm{HL}}$, respectively), but this does not hold for the
r50\%-ML model, and there is even a marked deterioration in the case of
10\%-ML.}

\section{Conclusions}
\label{sec:conc}

 We propose a procedure for building highly accurate PESs using KRR-based 
machine learning. Our approach employs structure-based sampling to ensure that 
machine learning is applied in the interpolation regime where it performs best. 
Data slicing in terms of the energy distribution is recommended for a balanced 
representation of the PES \red{for very small training sets}. Self-correction capabilities are introduced into the 
ML model by including additional ML layers. 

 In a pilot study, we explored the merits of our ML model using a recently 
published high-level \abinitio{} PES of CH$_3$Cl as an example. Several 
ML models were built and trained using training sets of different size and a 
different number of data slices, and their performance was assessed by 
comparisons with the original \abinitio{} energies at the grid points in 
the training set and the prediction set. Excellent agreement was found for the 
50\%-ML model, \red{which reproduces the PES with sub-wavenumber accuracy}.
 
 For \red{five} selected ML-based PESs, the vibrational energy levels of CH$_3$Cl 
were computed using variational TROVE calculations, in complete analogy to the 
published \abinitio{} work~\cite{15OwYuYa.CH3Cl}. The results clearly 
show that structure-based sampling produces more accurate ML 
models than random sampling. The structure-based sampling 50\%-ML model gives 
excellent agreement with the \abinitio{} reference data for the 
vibrational energy levels (RMSE of 0.04~\wn{} in the range up to 10,000~\wn{}). 
\red{The accuracy deteriorates slightly for the 25\%-ML model (RMSE of \red0.16~\wn{}) 
and for the s10\%-ML model (RMSE of 0.28~\wn{}, training set sampled from
the dataset sliced into three regions)
and quite strongly for the 10\%-ML model (RMSE of 1.75~\wn{}, training set sampled from
the unsliced dataset).}

 The evidence from the present pilot study suggests that the number (and 
computational cost) of electronic structure calculations in high-level 
\abinitio{} studies of rovibrational spectroscopy may be reduced by \red{up to 90\% (depending on the user needs and the initial grid size)}
by using structure-based sampling and self-correcting ML models, 
with minimal loss of accuracy. We expect that this will also hold for small 
molecules other than CH$_3$Cl (in the absence of obvious reasons to suspect 
anything else). Of course, this should be examined in future work, which should 
also aim at establishing standard protocols for such ML studies. Finally, given 
the fact that our ML model is available in analytic form, it seems worthwhile 
to explore whether it can be used directly in TROVE-type variational 
calculations (without an intermediate fit to a standard polynomial form).
 
\section{Outlook}
\label{sec:outlook}
The objective of this study was not to improve upon the existing \abinitio{} 
PES of \ce{CH3Cl} but to demonstrate how the computational cost of building 
such a PES can be substantially reduced by performing fewer \abinitio{} 
calculations and by interpolating efficiently with KRR-based ML. We have shown 
that this can be done by the following procedure.
\begin{enumerate}
 \item Generate a large and dense grid of deformed structures using established 
techniques (as outlined in Section~\ref{subsec:sampling} and 
Ref.~\citenum{15OwYuYa.CH3Cl}).
 \item Select points from this grid into the training set by using 
structure-based sampling.
 \item Calculate the energy for each point in the training set as accurately as 
possible using a high-level \abinitio{} method.
 \item Train the self-correcting ML model on the training set geometries and 
the high-level \abinitio{} energies.
 \item Predict the energies for the remaining grid points using self-correcting 
machine learning.
 \item Calculate rovibrational levels variationally using TROVE.
 \item Repeat Steps 2 to 6 by including more points from the grid into the 
training set using structure-based sampling until the calculated rovibrational 
levels converge.
\end{enumerate}
We plan to apply this procedure to generate accurate PESs for other small 
molecules. \red{During these studies we will also investigate additional
ways to reduce even further the computational cost of generating
accurate PESs by combining our procedure with the so-called
$\Delta$-ML approach\cite{Dral_Delta_ML_2015}. In
this approach, ML is trained on differences between high-level and 
less demanding but still reliable low-level \abinitio{} results; the $\Delta$-ML model is then
applied to correct low-level energies for
the remaining grid points (similar to previous work on
the water PES\cite{Bartok_GPwater_2013}). We anticipate that this approach will allow us to investigate larger systems such as organic molecules with several carbon atoms.}

\begin{acknowledgments}
PD thanks Georgios Gerogiokas for software performance related advice and 
Alexey Dral for discussions. \red{The authors thank the reviewers for their
valuable comments.}
\end{acknowledgments}

%

\end{document}